\documentclass[3p,times,twocolumn]{elsarticle}

\usepackage{ecrc}

\volume{00}

\firstpage{1}

\journalname{Nuclear Physics B Proceedings Supplement}

\runauth{}

\jid{nuphbp}

\jnltitlelogo{Nuclear Physics B Proceedings Supplement}

\usepackage{amssymb}

\usepackage[figuresright]{rotating}

\begin{document}

\begin{frontmatter}



\dochead{}

\title{Fermi bubbles as a source of cosmic rays above $10^{15}$ eV}


\author[HKU,LPI,NCU]{D.O. Chernyshov}
\author[HKU]{K.S. Cheng}
\author[LPI]{V.A. Dogiel}
\author[NCU]{C.M. Ko}

\address[HKU]{Department of Physics,
University of Hong Kong, Pokfulam Road, Hong Kong, China}
\address[LPI]{I.E.Tamm Theoretical Physics Division of P.N.Lebedev
Institute of Physics, Leninskii pr. 53, 119991 Moscow, Russia}
\address[NCU]{Institute of
Astronomy, National Central University, Jhongli, Taiwan}

\begin{abstract}

Fermi bubbles are giant gamma-ray structures extended north and south
of the Galactic center with characteristic sizes of order of 10 kpc recently
discovered by Fermi Large Area Telescope. Good correlation between radio
and gamma-ray emission in the region covered by Fermi bubbles implies the
presence of high-energy electrons in this region. Since it is relatively difficult
for relativistic electrons of this energy to travel all the way from the Galactic
sources toward Fermi bubbles one can assume that they accelerated in-situ.
The corresponding acceleration mechanism should also affect the distribution
of the relativistic protons in the Galaxy. Since protons have much larger
lifetimes the effect may even be observed near the Earth. In our model we
suggest that Fermi bubbles are created by acceleration of electrons on series
of shocks born due to periodic star accretions by supermassive black hole
Sgr A*. We propose that hadronic CR within the ''knee'' of the observed
CR spectrum are produced by Galactic supernova remnants distributed in
the Galactic disk. Reacceleration of these particles in the Fermi Bubble
produces CRs beyond the knee. This model provides a natural explanation
of the observed CR flux, spectral indexes, and matching of spectra at the
knee.
\end{abstract}


\end{frontmatter}


\section{Introduction}
Since their discovery by \citet{dob1} and \citet{meng} the giant gamma-ray structures also known as ''Fermi bubbles'' located above and below the Galactic center remain as one of the most attractive astrophysical events. Despite their nature is still enigmatic the location of these objects indicates their connection with past or present activity in the center of our Galaxy. Different models relate the bubbles to starburst activity \citep{crock}, single \citep{fuj13, guo12, meng, yang12} or multiple \citep{cheng11} energy release events on a central black hole.

Fermi bubbles also are observed in other wavelengths. In particular observations in microwave band show a very good correlation of so-called ''WMAP haze'' with gamma-ray emission \citep{ade13,fink}. There are some indications on the hot plasma inside the Fermi bubbles observed by a {\it ROSAT} as a narrow envelope with very sharp edges \citep{bl03}. This structure is explained as a fast wind with a velocity $u_w \sim 10^8$ cm/s driving a shock into the halo gas. However subsequent observations of Fermi bubbles edges by {\it Suzaku} did not find any evidences of a strong shock there \citep{suz_bubble}.

Since Fermi bubbles are very faint structures it is impossible to observe exactly the same phenomena in other galaxies. However some much more powerful objects with similar properties and probably similar nature are observed in some galaxies with active nuclei. For example even more giant structures are clearly seen in the direction of Cen-A in GHz radio \citep{feain11,junkes93}, GeV \citep{yang12a} and TeV \citep{aha09} gamma-ray ranges. Giant X-ray and radio lobes (bubbles) were found also  in the galaxies  NGC 3801 \citep{croston}, Mrk 6 \citep{mingo11} and Circinus Galaxy \citep{mingo12}.

Taking all facts together one can conclude that Fermi bubbles are indeed real structures. Since they enclose huge energy \citep{meng} and their volume is comparable to that of the Galactic disk Fermi bubbles have a potential to affect the distribution of the cosmic rays (CRs) in the Galaxy. In particular in our paper \citep{cheng12} we showed that Fermi bubbles may be responsible for the formation of the CR spectrum above the ''knee''. Below we will briefly recapitulate the major points of this model and also discuss some consequences.

\section{Reacceleration of the hadronic component of CR by Fermi bubbles}
It is generally accepted that supernova (SN) explosions in our Galaxy can provide enough energy to produce the observed total luminosity of the CR  \citep[see e.g.][]{ber90}. Moreover diffusive shock acceleration naturally explains power-law spectrum of CR \citep{bell78,krym} and taking into account propagation effects their spectral index. However many fundamental questions related to the assumption of SNRs being the sources of the Galactic CR are still open. One of them is the maximum energy of CR which can be estimated from the age of a typical SNR $T$ expanding with velocity of $u_{sh}$
\begin{equation}
E_{max} \sim Ze\beta_{sh}  B  T u_{sh}\,,
\label{EqEmax}
\end{equation}
where $\beta_{sh}=u_{sh}/c$, $B$ is the magnetic field strength at
the shock and the term $\mathcal{E}=\beta_{sh} B$ in this case can
be interpreted as an effective electric field.

For the parameters of the standard Galactic SNR and for reasonable values of the Galactic magnetic field the maximum energy of CR protons cannot exceed $10^{13} - 10^{14}$ eV. In more complex models outside quasi-linear approximation it was shown that the magnetic field at the shock can be amplified. As a result for the conservative set of parameters maximum energy reaches the value of about $10^{15}$ eV.

The important point of the CR spectrum is sudden steepening around $3\times 10^{15}$ eV which indicates on the change of the acceleration or propagation mechanism. Smooth attachment of the spectra above and below this energy and sharpness of the transition indicates that we are dealing with sole spectrum rather than with sum of two distinct components. The review of different models suggested to explain this phenomenon can be found in the original paper \citep{cheng12}.

In summary, it is generally agreed that SN shocks can only accelerate particles to energies $E<10^{15}$eV. However Fermi bubbles with age and size exceeding that of typical SNR by 4 orders of magnitude can easily accelerate particles to much higher energies. Using Eq. (\ref{EqEmax}) one can estimate that the bubbles have a potential to form the spectrum of CR in between $3\times 10^{15}$ eV and $10^{18}$ eV. We do not consider the possibility that Fermi bubbles can form a whole spectrum of CR both below and above $3\times 10^{15}$ eV. The reason for this is that the total power needed for the luminosity of CRs in our
Galaxy $L_{CR} \sim 10^{41}$ erg/s \citep{berezh1} and the average energy release toward the Fermi bubbles is also $10^{41}$ erg/s \citep{cheng11}. To explain the origin of CR by acceleration in the Fermi bubbles one require the acceleration efficiency to be close to 100\% which is unlikely. On the other hand the sudden change of the slope of the spectrum around the ''knee'' can be naturally explained due to change of the acceleration properties. Thus our model can be described in the following way: SNR in the disk accelerate particles with power-law distribution up to energies of $3\times 10^{15}$ eV and Fermi bubbles further re-accelerate this particles up to $10^{18}$ eV.

In the model of Fermi bubbles origin we suggested that the activity responsible for the formation of the bubbles was due to stellar capture and tidal disruption by a central black hole \citep{cheng11, cheng12}. The average time between two successive captures in the Galaxy is between $10^4$ yrs and $10^5$ yrs \citep{syer1999}. Thus the activity is periodic and characteristic period between two event is shorter than the characteristic lifetime of the Fermi bubbles. Periodic energy releases in the Galactic center should form series of shock propagating through the halo. We note that from the numerical simulations it appears that only very powerful events expected in the case of capture of massive stars can form shocks. Thus the amount of shocks inside the Fermi bubbles should not be very large.

\begin{figure}[h]
\begin{center}
\includegraphics[width=0.4\textwidth]{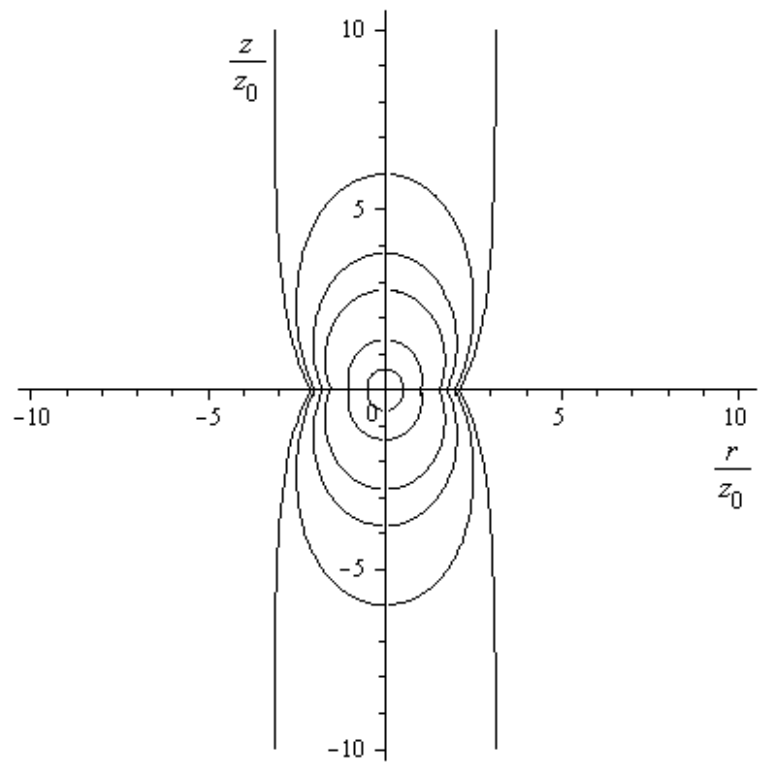}
\end{center}
\caption{The bubble
multi-shock structure. Five representative shocks are shown and each shock is injected into the halo in a time interval of $2\times 10^6$yr. Explosion energy was taken as $W = 3\times 10^{52}$ erg, $\rho_0 =1.6\times 10^{-24}$g/cm$^3$ and $z_0=1$ kpc are used for illustration purpose.}\label{fig:komp}
\end{figure}

In the exponential atmosphere of the halo with the scale $z_0$, i.e. $\rho(z)= \rho_0 \exp(-z/z_0)$,  an  analytic  solution of shock propagation was obtained by \citet{komp}. Figure \ref{fig:komp} illustrates the process of shock propagation. This figure is meant to be illustrative only since it is valid only for cold homogeneous atmosphere with exponential density profile. Environment of the Galactic halo is expected to be more complex especially if it is affected by strong outflows from the Galactic center. Thus in reality the distribution of shocks should be far more complicated. However this simplified picture is able to explain the shape of the bubbles as well as their characteristic size.

The separation between shocks can be estimates from their velocity $u$ and characteristic period of captures $\tau_{cap}$ as
\begin{equation}
l_{sh}=\tau_{cap}u=30(\tau_{cap}/3\times 10^4\mbox{ yr})(u/10^8\mbox{ cm/s})\mbox{ pc.}
\end{equation}
However the exact separation between two consecutive shocks depends on the actual time separation of two consecutive capture events and their energy releases. There is another important spatial scale which characterizes processes of particle acceleration by a single shock ,i.e. the acceleration length scale in a single shock $l_D \sim D/u$, where $D$ is the spatial diffusion coefficient
near a shock whose value depends on particle interaction with small scale magnetic fluctuations and $u$ is the shock velocity.

The problem of particle acceleration in conditions of supersonic
turbulence (multi-shock structure) was extensively analyzed before.
In series of papers by \citet{byk90, byk92,byk01} as applied to acceleration
processes in OB associations,which is quite similar to the structure of the Bubble, they introduce a nondimensional parameter characterizing acceleration regime as
\begin{equation}
\psi=\frac{l_{sh}}{l_D}=\frac{ul_{sh}}{D}\sim \frac{ul_{sh}}{cr_L}\,.
\end{equation}
The corresponding energy $E_1$ that separates different regimes can be estimated from the condition $\psi\sim 1$ or $l_D (E_1)\sim l_{sh}$ which for the conditions of the Fermi bubble is
\begin{eqnarray}
E_1 &&=eBl_{sh}u/c \nonumber\\
    &&= 10^{15}
\left(\frac{B}{5~\mu G}\right)
\left(\frac{l_{sh}}{30~\mbox{pc}}\right)
\left(\frac{u}{10^8~\mbox{cm/s}}\right)~\mbox{eV.} \label{e1}
\end{eqnarray}
In the case of  $\psi\gg 1$ or $l_{sh} \gg l_D$ analyzed in
\citet{byk90,byk92}, there is a combined effect of a fast particle
acceleration by a single shock, which generates the spectrum
$E^{-2}$   and relatively slow transformation of this
spectrum due to interaction with other shocks (stochastic Fermi
acceleration) into a hard $E^{-1}$ spectrum in the intershock
medium at relatively low energies. However it is unclear if such slow transformation can be completed within the life time of the shocks in the Bubble. Furthermore the hard $E^{-1}$ spectrum requires significantly higher power to be formed in comparison with softer spectrum with the slope of $E^{-2}$. Thus the transformation of the spectrum requires the power which significantly exceeds $10^{41}$ erg/s which the bubbles as we mentioned above can not supply. It is reasonable to assume that Fermi bubbles do not affect the spectrum of CR below $E_1$. This statement is in agreement with our initial assumption that SNRs are the major contributors for CRs with energies $E<10^{15}$eV and the exact particle spectrum generated from the Bubble is unimportant in the energy range of $E<10^{15}\rm eV$.

For  $\psi\ll 1$ or $l_{sh}\ll l_D$ the acceleration regime  changes to a
pure stochastic acceleration by a supersonic turbulence. In the
stationary case the equation for accelerated CRs can be presented in the form \citep{byk90}
\begin{equation}
\frac{\partial}{\partial z}D(\rho)\frac{\partial f}{\partial
z}+\frac{1}{\rho}\frac{\partial}{\partial
\rho}D(\rho)\rho\frac{\partial f}{\partial
\rho}+\frac{1}{p^2}\frac{\partial}{\partial p}\kappa(\rho)
p^2\frac{\partial f}{\partial p}=0 \label{dif_bubble}
\end{equation}
where $\rho$ and $z$ are the cylindrical spatial coordinates, $p$
is the particle momentum, $D(\rho)$ is the spatial diffusion
coefficient and $\kappa$ is the momentum diffusion coefficient.

Proton acceleration in the Bubble depends sensitively on the
acceleration parameters and structure of the Bubble. In the
following we present a detailed analysis. We present the bubble
region as a cylinder extending above and below the Galactic plane
from $z=0$ to $z=\pm H $ with a radius $\rho=\rho_B$. As
the boundary conditions we put the density of particles $f$
equaled zero at the Galactic halo surface
\begin{equation}
f|_\Sigma =0~~\mbox{at}~~\rho=\rho_G,~~\mbox{and}~~z=H
\end{equation}
The diffusion coefficients inside and outside the bubble are supposed
to be different
\begin{eqnarray}
&&D(\rho)=D_B\theta(\rho_B-\rho)+D_G\theta(\rho-\rho_B)\nonumber \\
&&\kappa=\kappa_B\theta(\rho_B-\rho)\Theta(E-E_1)
\label{spatialdiff}
\end{eqnarray}
 where $D_B=Lc/3$ is the coefficient inside the bubble due to
 interactions with a supersonic turbulence and $D_G$ is the average
 diffusion coefficient in the Galaxy defined e.g. in
 \citet{ber90}. The momentum diffusion coefficient is
 $\kappa_B=u^2/D_D$.
The momentum dependence of $f$ can be presented by a power-law
function, $f(p)\propto p^{-\gamma}$, where $\gamma$ should be
determined from Eq.(\ref{dif_bubble}) and approximately is
\begin{equation}
\gamma\approx\frac{3}{2}+\sqrt{\frac{9}{4}+\pi^2\frac{D}{\rho_B^2\kappa}} =
\frac{3}{2}+\sqrt{\frac{9}{4}+\frac{\pi^2D_B^2}{u^2\rho_B^2}}\,.
\end{equation}
Thus a proper choice of $D_B$ allows to form a power-law distribution with a certain value of a spectral index.

To consolidate this idea, we work out a concrete numerical model.
Essentially, we solve the stationary state CR
transport equation~(\ref{dif_bubble}) in our Galaxy with
two Fermi Bubbles (one on each side of the Galactic plane).
We modeled our Galactic halo as a cylinder of
radius $\rho_G=20$ kpc, and the top and bottom at $\pm 10$ kpc
from the mid-plane. Each Fermi Bubble is also a cylinder of the
same height $\pm 10$ kpc, but with a radius $\rho_B=3$ kpc.
The spatial diffusion coefficient are different inside and outside the bubble as
described by Equation~(\ref{spatialdiff}).

Since we expect the average separation between shock in the Fermi bubbles to be of order of 100 pc \citep{cheng12} we consider a constant spatial diffusion
coefficient and adopt $D_B=2.08\times 10^{30}$ cm$^2$ s$^{-1}$.
Outside the bubble, we take into account the energy (or momentum)
dependence of the spatial diffusion coefficient and adopt
$D_G=D_0(pc/4\,{\rm GeV})^{0.6}$, $D_0=6.2\times 10^{28}$ cm$^2$ s$^{-1}$
\citep[cf.][]{acker}.

We assume that there is little or no stochastic acceleration outside the bubble (see Eq. (\ref{spatialdiff})), and adopt $\kappa_B H^2/D_B=1.9$ (i.e., $\kappa_B=4.4\times 10^{-15}$ s$^{-1}$ or the corresponding acceleration time scale is $7.6$ Myr).

For Galactic SNRs adopt the distribution suggested by \citet{stecker}
and modified it with a Gaussian thickness profile
\begin{equation}\label{SNRdist}
Q_{\rm SNR}(\rho,z)\propto \left({\rho\over R_\odot}\right)^{1.2}
\exp\left(-\,{3.22\rho\over R_\odot}\right)\exp\left(-{z^2\over h^2}\right)\,,
\end{equation}
here we take $h=100$ pc, $R_\odot=8$ kpc.
We adopt the idea that SNRs inject energetic particles in the form of a power law with a high-energy cutoff at $p_{\rm max} c\approx 3\times 10^{15}$ eV. Therefore, together with the SNR distribution (Equation~(\ref{SNRdist})), the source function is
\begin{eqnarray}
Q(\rho,z,p)&=Q_0 \left({p\over p_{\rm max}}\right)^{-\mu}\exp\left(-\,{p\over p_{\rm max}}\right)
\left({\rho\over R_\odot}\right)^{1.2} \nonumber\\
&\times \exp\left(-\,{3.22\rho\over R_\odot}\right)\exp\left(-{z^2\over h^2}\right) \,.
\label{eq:Q}
\end{eqnarray}

Finally, the appropriate boundary conditions for the momentum
coordinate are
\begin{equation}
\left.{p\over f}{\partial f\over\partial p}\right|_{p=p_{\rm low}}= -4.7\,,\quad
f|_{p=p_{\rm up}} = 0\,,
\end{equation}
where the energy of the lower momentum boundary is $p_{\rm low}c = 10^{12}$ eV, and
the upper momentum boundary is $p_{\rm up}c = 3\times 10^{18}$ eV.
The condition at the lower momentum ensures that the spectral index matches that of
low-energy CRs (say $E<10^{12}$ eV).

The spatial boundary conditions are
\begin{equation}
\left.{\partial f\over\partial\rho}\right|_{\rho=0} = 0\,,\quad f|_{\rho=\rho_G} = 0\,,
\end{equation}
and
\begin{equation}
\left.{\partial f\over\partial z}\right|_{z=0} = 0\,,\quad f|_{z=\pm H} = 0\,.
\end{equation}

\begin{figure}[ht]
\begin{center}
\includegraphics[width=0.4\textwidth]{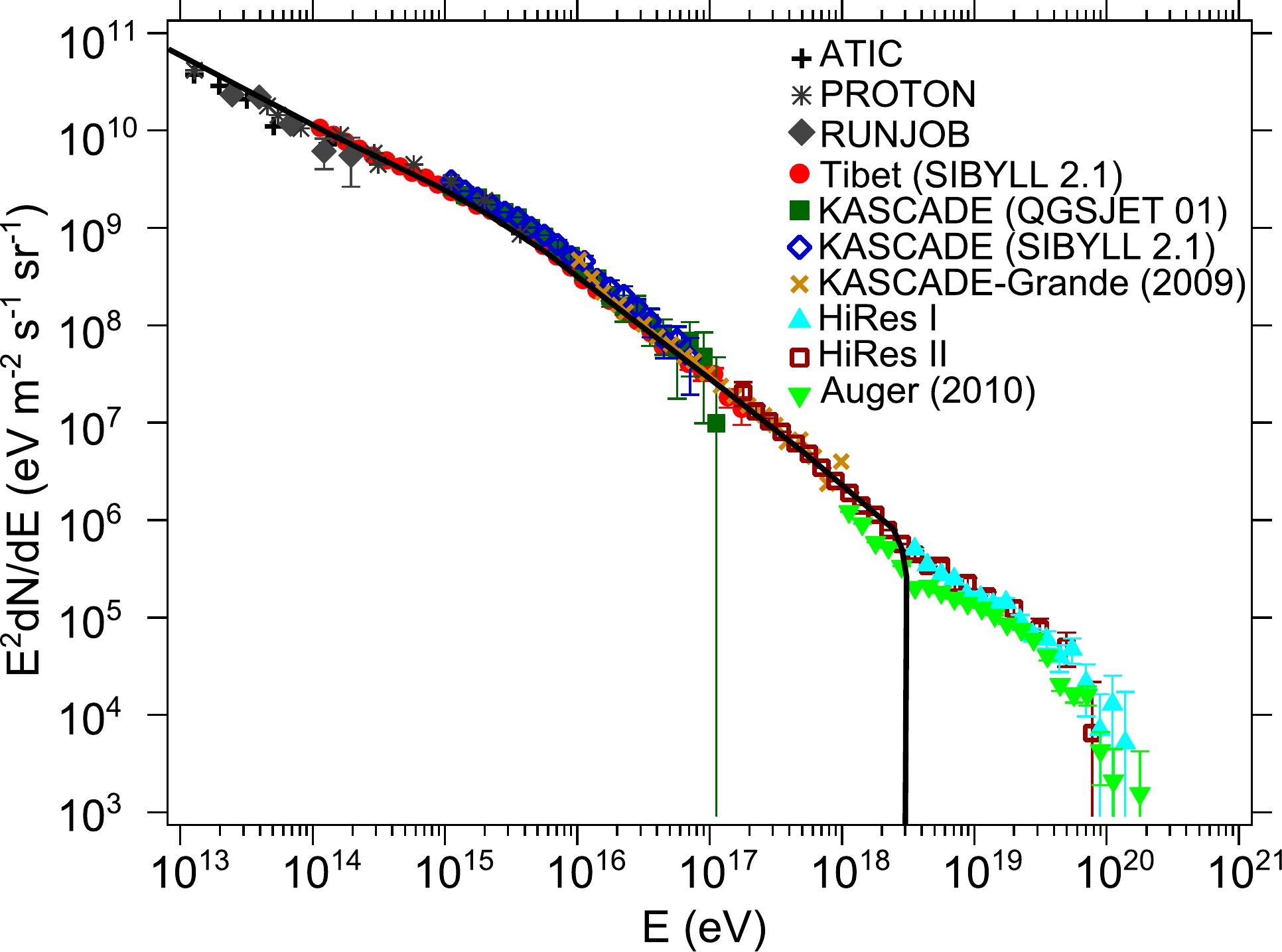}
\end{center}
\caption{CR spectrum at the Earth as a combination of the SNR contribution (in the Galactic disk) and the stochastic acceleration in the Fermi Bubbles.
The black solid line is the spectrum from our numerical model.
For the references to the experimental data see the original paper \citep{cheng12}.
In this model, $D_B=2.08\times 10^{30}$ cm$^2$ s$^{-1}$ inside the bubbles
and $D_G=6.2\times 10^{28} (pc/4\,{\rm GeV})^{0.6}$ cm$^2$ s$^{-1}$ outside,
$\kappa_B H^2/D_B=1.9$, and the injection spectrum from SNR $\mu=4.35$ (see Eq. (\ref{eq:Q})).
}
\label{fig:dif+fermi2}
\end{figure}

The spectrum evaluated at Earth's position is the solid line shown
in Figure~\ref{fig:dif+fermi2}.
The model fits the data reasonably well and it is not coincident that the
spectra join smoothly at the knee. The value of the parameter $E_1$ based on Eq. (\ref{e1}) is not reliable. To check how the result depends on this parameter we estimated the spectrum for different values of $E_1$ (see Eq. (\ref{spatialdiff})). It turns out that the shape of the spectrum does not depend on $E_1$ as long as $E_1 < 3\times 10^{15}$ eV.

We should note that in the estimations above we assumed that CR spectrum consist only of protons. This assumptions is for illustrative purposes only. To obtain the spectrum of different elements one should note that propagation and acceleration parameters depend only on rigidity. Thus the spectra of all elements should have the same shape but should be shifted along the energy axis in accordance to their charge $Z$ \citep[see e.g. Fig. 5 in][for the details]{kamp}. Due to the same shape of the spectrum one should expect several ''knees'' in the spectrum corresponding to different elements at the energies of $3Z\times 10^{15}$ eV. According to \citet{apel} there is a knee-like structure in the heavy component of CR spectrum at about $8\times 10^{16}$ eV (apparently corresponding to iron) that strengthen this idea. 

In this case spectrum of particles above the knee should be slightly softer than the one estimated by us for proton-only case. But as one can see from Eq. (\ref{e1}) the value of the spectral index can be easily adjusted by a proper choice of the acceleration parameters.

\section{Reacceleration of the leptonic component of CR by Fermi bubbles}
Propagation of charged particles in the Galaxy is mainly determined by their energy losses. Since relativistic protons spend most of their time in the Galactic halo their energy losses are insignificant and can be neglected. The same conclusion is also valid for electrons with energies below hundred GeV. Indeed in the Galactic halo environment the density of soft photons and magnetic field strength are low allowing electrons to diffuse to more than 10 kpc before losing energy due to synchrotron and inverse-compton losses.

Since propagation of electrons is similar to that of protons one can expect that Fermi bubbles should also affect the population of CR electrons near the Earth. The effect of reacceleration should appear as a ''bump'' in the spectrum of electrons at energies of about 10-100 GeV. However even before the influence of the Fermi bubbles can be detected in the spectrum of CR electrons the effect of the reacceleration should be visible in the gamma-ray background. This phenomenon should appear as a gamma-ray image of the Fermi bubbles.

We note that in the frame of multi-shock model \citep{cheng11} with the parameters described in the previous section electrons do not ''feel'' multiple shocks and their acceleration should be described as ordinary diffusive shock acceleration (since $\psi\gg 1$, see the previous section). Even if one assume that the separation between shocks is very small and $\psi$ becomes less than unity even for electrons with energy less than 1 TeV the spectrum of re-accelerated electrons would be too soft and thus the overall spectrum of the electrons would not be affected.

In that sense reacceleration of electron responsible for gamma-ray emission from Fermi bubbles and reacceleration of CR hadrons have generally speaking different nature. However we think that it is important to mention the possible difficulties which can appear in all models of Fermi bubbles origin based on direct \citep[e.g.][]{cheng11} or indirect \citep[e.g. in][via bow shock]{guo12, yang12} acceleration of electrons.

The reacceleration process can be roughly described as a smooth attachment of a power-law tail to the background distribution function i.e. the reaccelerated spectrum looks like
\begin{equation}
f_r(E) = f_0(E)\Theta(E_c-E) + 
f_0(E_c)\left(\frac{E}{E_c}\right)^{-\delta_1}\Theta(E-E_c)\,,
\label{reacc1}
\end{equation}
where $f_0(E)$ is the initial spectrum of particles before the reacceleration and $E_c$ is determined from the continuity of the particle flow in the energy space. In most cases $E_c$ can be estimated from the continuity of the derivative $\partial f / \partial E$. In the case of Galactic CR electrons the spectrum $f_0(E)$ have a sharp spectral brake near 2.2 GeV so it is reasonable to assume that $E_c \approx 2$ GeV.

In order to estimate the gamma-ray emission from the Fermi bubbles one should integrate the spectrum of electrons over the line of sight. Assume for simplicity that all reaccelerated electrons reside in the shell of thickness $\Delta x$ and introduce an ''averaged'' spectrum:
\begin{equation}
<f_r(E)> = \frac{1}{\Delta x}\int f_r(r,E)dl \,,
\end{equation}
where $l$ is the distance along line of sight. Since electrons at GeV energies are weakly affected by losses and the distribution of them within the Fermi bubble is almost constant
\begin{equation}
<f_r(E)> = f_0(E)\Theta(E_c-E) + 
f_0(E_c)\left(\frac{E}{E_c}\right)^{-\delta}\Theta(E-E_c) \,,
\end{equation} 
where $\delta$ is the average spectral index of the reaccelerated particles along the line of sight. Generally speaking $\delta \neq \delta_1$ from Eq. (\ref{reacc1}) due to effect of energy losses.  The value of $\delta$ may weakly depend on energy but in this analysis we assume it constant for the sake of simplicity.

\begin{figure}[ht]
\begin{center}
\includegraphics[width=0.45\textwidth]{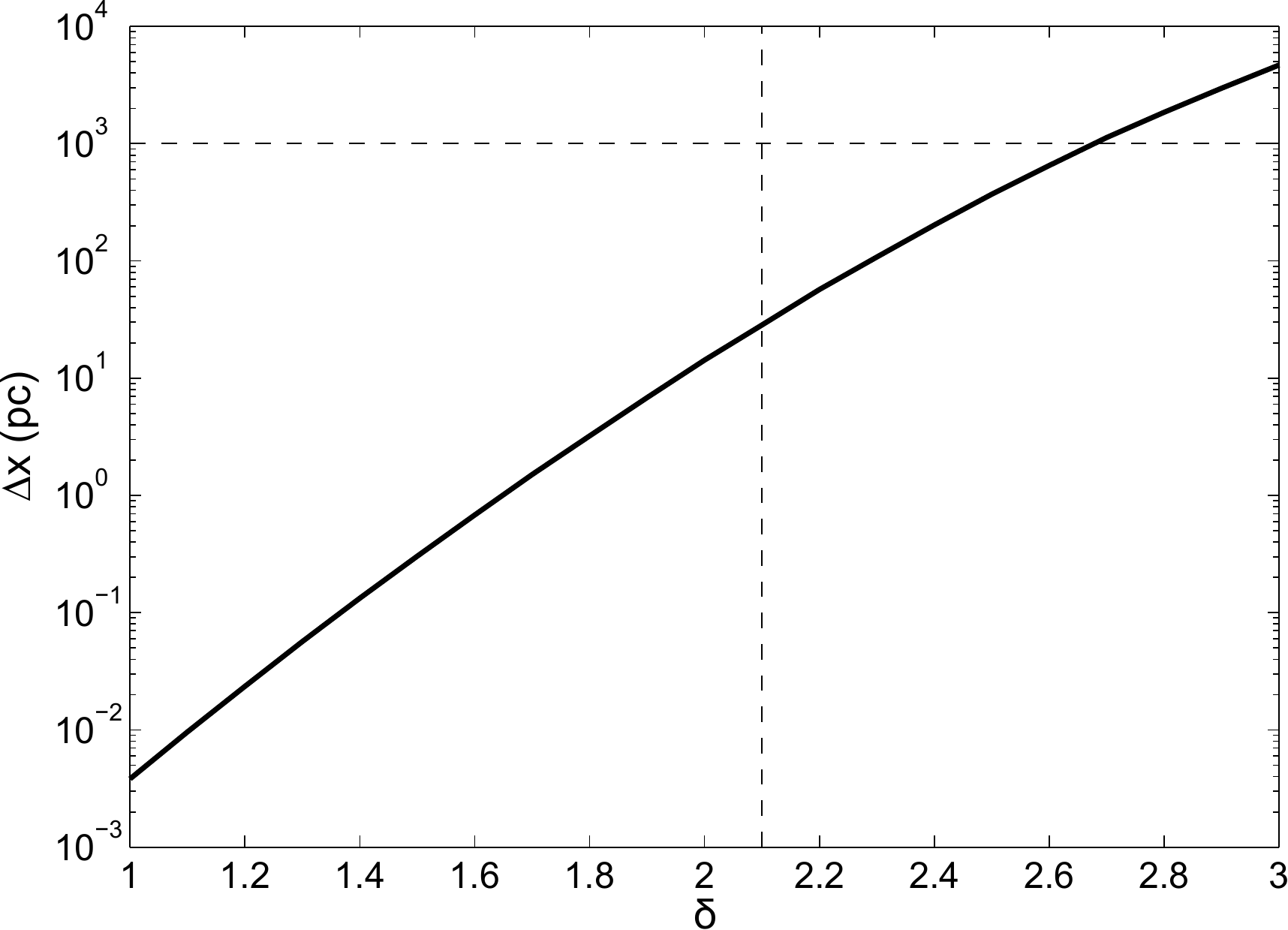}
\end{center}
\caption{Thickness of the bubble walls $\Delta x$ necessary to produce observed gamma-ray flux (thick line) depending on the spectral index of relativistic electrons inside the bubble $\delta$. Vertical dashed line marks the expected spectrum of electrons based on radio data \citep{ade13}. Horizontal dashed line marks the expected thickness based on the spatial structure \citep{meng}. \label{fig:DDx}}
\end{figure}

The intensity of the inverse Compton emission can be estimated in the following way
\begin{equation}
I_{IC} = \frac{\Delta x}{4\pi}\int\limits_{E_{IC}^{(min)}}^{E_{IC}^{(max)}}
<f_r(E)>\left(\frac{dE}{dt}\right)_{IC} dE \,,
\end{equation}
where $E_{IC}^{(min)}$ and $E_{IC}^{(max)}$ is energy of the electrons corresponding to maximum and minimum energy of gamma-ray photons accordingly and $(dE/dt)_{IC}$ are inverse Compton energy losses. Using for simplicity non-relativistic approximation $(dE/dt)_{IC} = \beta_{IC}E^2$ and assuming that $\delta < 3$ one can obtain that
\begin{equation}
\Delta x = I_{IC}\frac{4\pi(3-\delta)}{\beta_{IC}f_0(E_c)} \left(\frac{E_{IC}^{(max)}}{E_c}\right)^{\delta-3}E_c^{-3} \,.
\label{DeltaX}
\end{equation}
According to the one of the publicly available GALPROP simulations
$S_SZ_8R_{20}T_{\inf}C_2$ \citep{acker} the density of electrons at 2 GeV is
$f_0(E_c) =  8.7\times 10^{-16}$ cm$^{-3}$MeV$^{-1}$. The cut-off in the gamma-ray spectrum requires $E_{IC}^{(max)} \approx 0.3$ TeV \citep{meng} then
\begin{eqnarray}
\Delta x &= 2.6\mbox{ kpc} \times 
\left(\frac{I_{IC}}
{4\times 10^{-4}~\mbox{MeV}\cdot\mbox{cm}^{-2}\mbox{s}^{-1}\mbox{sr}^{-1}}
\right) \nonumber \\
&\times\left(\frac{w_{ph}}{1~\mbox{eV}\cdot\mbox{cm}^{-3}}\right)^{-1}
\nonumber \\
&\times\left(
\frac{f_0(E_c)}{8.7\times 10^{-16}~\mbox{cm}^{-3}\mbox{MeV}^{-1}}
\right)^{-1} \nonumber \\
&\times\left(\frac{E_c}{2~\mbox{GeV}}\right)^{-3} \times
\left(\frac{E_{IC}^{(max)}}{E_c}\right)^{\delta-3}\,,
\label{thickness_f}
\end{eqnarray}
where $w_{ph}$ is the energy density of the soft photons.  If we assume that $\delta = 2.1$ as it is required by radio observations \citep{ade13} we obtain that the thickness of FERMI bubble walls is only $\Delta x = 30$ pc. This value is much smaller than the value of 1-2 kpc required to reproduce the spatial shape of the bubble and also require an extremely small spatial diffusion coefficient to confine relativistic electrons in a thin shell. A finer $\Delta x(\delta)$ relations based on the correct expressions for the gamma-ray and radio emission together with some restriction from observation data are presented in Fig. \ref{fig:DDx}.

One can conclude that reacceleration process overproduces relativistic electrons. This effect does not depend on the nature of the reacceleration mechanism as long as it produces power-law spectrum of electrons. To avoid this problem it is necessary to introduce another kind of losses which should be effective in GeV energy range. One of the possible solutions is introduction of adiabatic losses. Indeed if the plasma flow in the Galactic halo in the vicinity of Fermi bubbles is non-uniform particles start to lose energy with a rate
\begin{equation}
\frac{dE}{dt} = \frac{E}{3}~\nabla\cdot\bf u\,,
\end{equation}
where $\bf u$ is the flow velocity. According to \citet{bloe} the velocity of the Galactic wind can be described as linear function of the altitude $u(z)=3\upsilon_0 z$ and the gradient value $\upsilon_0$ can be as high $\upsilon_0\simeq 10^{-15}$s$^{-1}$. With this rate adiabatic losses become essential in GeV energy range and are able to significantly reduce the value of $f(E_0)$ in (\ref{thickness_f}) and soften the restrictions on $\Delta x$ and $\delta$.

\section{Conclusion}
We suggested a model of CR origin in energy range $3\times 10^{15}$ eV $< E < 10^{18}$ eV. In the model we assumed that CR with energies below $3\times 10^{15}$ eV are produced by SNR in the Galactic disk. This particles diffuse out to the Galactic halo and then are reaccelerated in the Fermi bubbles. For a proper choice of the acceleration parameters it is possible to reproduce the right value of the spectral index in energy range above the ''knee''. The spectrum below the ''knee'' remains undisturbed. This model also naturally explains the smooth attachment of the spectra near the ''knee''. It also explains the change of chemical composition of the CR at high energies if the rigidity dependence of the acceleration and propagation is taken into account.

We also estimated the possible effect of the reacceleration of the CR electrons in the Fermi bubbles. We showed the reacceleration process is very effective and it overproduces the expected gamma-ray flux from the Fermi bubbles. This effect does not depend on the reacceleration mechanism as long as it produces power-law spectrum of electrons. However this restriction can be softened by introduction of the adiabatic losses caused by non-uniformity of the plasma flow inside the Fermi bubbles.

\section*{Acknowledgments}
The authors would like to thank the unknown referee for his careful reading of the text and some suggestions useful for a future development of the model.
DOC is supported in parts by the RFFI grant 12-02-31648, the LPI Educational-Scientific Complex and Dynasty Foundation. DOC and VAD acknowledge support from the RFFI grant 12-02-00005. KSC is supported by the GRF Grants of the Government of the Hong Kong SAR under HKU 7010/13P. CMK is supported, in part, by the Taiwan National Science Council Grant NSC 102-2112-M-008-019-MY3.




\end{document}